# Emission of Gravitational Waves from a Magnetohydrodynamic Dynamo at the Center of the Sun

F. Winterberg

University of Nevada, Reno

Abstract

The failure of the laser-interferometer gravitational wave antennas to measure the tiny changes of lengths many orders of magnitude smaller than the diameter of a proton raises the question of whether the reason for this failure is a large gravitational wave background noise, and if so, where this background noise is coming from. It is conjectured that it comes from gravitational waves emitted from a magnetohydrodynamic dynamo in the center of the sun, with the large magnetic field from this dynamo shielded by thermomagnetic currents in the tachocline. Using the moon as a large Weber bar, these gravitational waves could possibly be detected by the Poisson diffraction into the center of the lunar shadow during a total solar eclipse.

1. Introduction

It was suggested by the author a long time ago [1], that there should be a cosmic gravitational wave background radiation as there is a cosmic electromagnetic background radiation, and that the gravitational background radiation might have made an observable imprint on the electromagnetic background radiation. A gravitational wave can be detected by a departure of the metric tensor from its Galilean value, changing the distance dx between two fixed positions by an amount set equal to $h$dx, where for the LIGO (laser interferometer gravitational wave observatory) h can be as small as $h \simeq 10^{-22}$. However, in spite of worldwide efforts to detect a gravitational wave signal, none has so far been seen. But if there is a large stochastic gravitational wave background, where $h$ is larger by orders of magnitude, and in a frequency range outside the accessibility of a LIGO antenna, this might obfuscate the observation of a much smaller non-stochastic gravitational wave event. This raises the question for the origin of this radiation. It is the conjecture made in this communication that it is the center of the sun from which large quantities of low-frequency gravitational waves are emitted, not accessible with a LIGO antenna. Following a conjecture of Alfvén [2] that in the center of the sun is a large magnetized plasma vortex torus, to explain the occurrence and the cycle of sunspots, which is a simple model for a magnetohydrodynamic dynamo in the center of the sun driven by thermonuclear reactions there taking place, it is here suggested that there are at least three such toruses, forming a Borromean ring, with a vanishing angular momentum to keep the sun flat. For a solar dynamo driven by the heat released from the thermonuclear reactions in the sun, a magnetic field as large as $B \simeq 10^8$ Gauss can be reached. Because such a large field is not observed at the surface of the sun, it must be shielded. This is possible by the large radial temperature gradient in the tachocline at the boundary of the thermonuclear reaction zone, leading to the generation of large thermomagnetic currents by the Nernst effect in this zone, large enough to shield the high magnetic field of the solar dynamo. By the large mass motion of the dynamo a flux of gravitational radiation is emitted, which in reaching the earth could lead to an $h$-value as large as $h \sim 10^{-7}$.

Using the moon with its large Q-value as a large Weber bar, the gravitational radiation can by Poisson scattering be focused into the center of the lunar shadow during a total solar eclipse where it may explain the strange Allais effect [3,4], observed during the eclipse on Foucault pendulums. Pendulums are very sensitive against mechanical disturbances, and for that reason they are used for the suspension of the laser reflectors in LIGO gravitational wave antennas, and for the same reason should be ideal for the observation of the Allais effect.

2. Magnetohydrodynamic Dynamo in the Center of the Sun

A magnetohydrodynamic dynamo is ruled, first, by the equation [5]

$$\frac{\partial \mathbf{B}}{\partial t} = \frac{c^2}{4\pi\sigma} \nabla^2 \mathbf{B} + curl \mathbf{v} \times \mathbf{B} \qquad (1)$$

where **B** is the magnetic field, **v** the fluid flow velocity, and σ the electrical conductivity of the fluid, second, by the continuity equation

$$\frac{\partial \rho}{\partial t} + div \rho \mathbf{v} = 0 \qquad (2)$$

where ρ is the density of the fluid, and third, by the Euler equation in the presence of a magnetic field, which for a frictionless fluid has the form

$$\frac{\partial \mathbf{v}}{\partial t} = -\frac{1}{\rho}\nabla p - \frac{1}{2}\nabla \mathbf{v}^2 + \mathbf{v} \times curl\mathbf{v} + \frac{1}{\rho c}\mathbf{j} \times \mathbf{B} \qquad (3)$$

where **j** is the electrical current density. A solution of these coupled non-linear equations is not possible in closed form, but some very general conclusions can be drawn. If the magnetic Reynolds number

$$\text{Re m} = \frac{4\pi\sigma Lv}{c^2} \gg 1 \qquad (4)$$

one can in (1) neglect the first term on the r.h.s. In the center of the sun the temperature is $T \simeq 1.6 \times 10^7 K$, with the conductivity $\sigma \simeq 2 \times 10^{18} s^{-1}$. Taking for the convection velocity for **v** in the absence of a magnetic field the value given by Schwarzschild [6], v≲10⁴cm/s, and for L≃10¹⁰cm the diameter of the thermonuclear reaction zone inside the sun one finds that Rem≃10¹²≫1. One can therefore replace (1) by

$$\frac{\partial \mathbf{B}}{\partial t} = curl\mathbf{v} \times \mathbf{B} \qquad (5)$$

Putting

$$\mathbf{v} = \boldsymbol{\omega} \times \mathbf{r} \qquad (6)$$

where **ω** = (1/2)curl**v**, and with (4π/c)**j**=curl**B**, one thus has for (3)

$$\frac{\partial \mathbf{v}}{\partial t} = -\frac{1}{\rho}\nabla p - \omega^2 \mathbf{r} - 2\boldsymbol{\omega} \times \mathbf{v} - \frac{1}{4\pi\rho}\mathbf{B} \times curl\mathbf{B} \qquad (7)$$

where the magnetohydrodynamic instabilities arise from the last term on the r.h.s.

If $(1/2)\rho v^2 \gg B^2/8\pi$ the fluid stagnation pressure $(1/2)\rho v^2$ overwhelms the magnetic pressure forces $B^2/8\pi$, and the magnetic lines of force align themselves with the streamlines of the fluid flow. If likewise the electric current flow lines **j** align themselves with **ω**, one has **j**⊥**B,** and **ω**⊥**v**. With **ω**=(1/2)curl**v** and **j**=(c/4π)curl**B** one can write for $(1/2)\rho v^2 \gg B^2/8\pi$

$$|\mathbf{v} \times curl \mathbf{v}| \gg \left|\frac{\mathbf{B} \times curl \mathbf{B}}{4\pi\rho}\right| \tag{8}$$

or that

$$v > v_A \tag{9}$$

where

$$v_A = \frac{B}{\sqrt{4\pi\rho}} \tag{10}$$

is the Alfvén velocity. Initially, one may set in equation (7) for |**ω**| the value for the slowly rotating sun, but with the build-up of the magnetic field by (5), **B** will eventually reach the value $B = \sqrt{4\pi\rho}v$, where $v_A = v$. In approaching this magnetic field strength, the magnetic pressure forces begin to distort the fluid flow which becomes unstable. In a plasma this leads to the formation of unstable pinch current discharges, where $p = B^2/8\pi$. For a hydrogen plasma of temperature T and particle number density n this leads to the Bennett equation

$$B = \sqrt{16\pi nkT} \tag{11}$$

The pinch instability which can be seen as the breakdown of the plasma into electric current filaments, and determined by the **B**×curl**B** term, with the **v**×curl**v** term is responsible for the breakdown of the plasma in vortex filaments. But while the breakdown into current filaments is unstable, the opposite is true for the breakdown into vortex filaments. This can be seen as follows: Outside a linear pinch discharge one has curl**B**=0, and outside a linear vortex filament curl**v**=0. Because of curl**B**=0, the magnetic field strength gets larger with a decreasing distance from the center of curvature of magnetic field lines of force. For curl**v**=0, the same is true for the velocity of a vortex line. But whereas in the pinch discharge a larger magnetic field means a larger magnetic pressure, a larger fluid velocity means a smaller pressure by virtue of Bernoulli's theorem. Therefore, whereas a pinch column is unstable with regard to its bending, the opposite is true for a line vortex. This suggests that a pinch column places itself inside a linear vortex filament.

What is true for the m=0 kink pinch instability is also true for the m=0 sausage instability by the conservation of circulation

$$Z = \oint \mathbf{v} \cdot d\mathbf{r} = const. \tag{12}$$

And because of the centrifugal force, the vortex also stabilizes the plasma against the Rayleigh-Taylor instability.

The magnetohydrodynamic dynamo in the sun is driven by thermonuclear reactions. In the presence of a magnetic field this leads to large plasma velocities magnifying the magnetic field up to the value given by (11), where the Alfvén velocity and plasma velocity are about equal to the thermal proton velocity $v_o \simeq \sqrt{kT/M}$. At T≃1.6×10$^7$K, the temperature in the solar core and M the mass of a proton, one has $v_o = 5.3 \times 10^7$ cm/s, and for the density in the center of the sun $\rho = nM = 1.5 \times 10^2$ g/cm³ ($n = 9 \times 10^{25}$/cm³). Inserting into (11) one finds that $B \approx 3 \times 10^8$ G, much larger than the magnetic dipole field of the sun which is on the order of 1 Gauss. It is conjectured that the mass motion of the dynamo in the center of the sun is a source of gravitational radiation, responsible for the Allais effect, but this strong magnetic field must be shielded from leaking out. This is possible by thermomagnetic currents in the tachocline, separating the region where the fusion reactions take place from the outer region of the sun.

The toroidal mass flow configuration in the center of the sun proposed by Alfvén [2], with the axis of the torus, was aligned with the axis of the rotating sun. In the context of the very large mass motion proposed here and to conserve angular momentum, the dynamo must have at least two such tori. However, since no sizable oblateness of the sun is observed, which would be caused by a large centrifugal force, more than one torus is needed to make the centrifugal force equal in all directions. The most simple topological configuration having this property is the 3-vortex Borromean configuration, which also minimizes the volume it occupies [7].

3. Emission of Gravitational Waves from the Core of the Sun

According to a theorem by Birkhoff, a completely spherical symmetric mass flow pattern cannot lead to the emission of gravitational radiation, but this theorem does not apply to a hydromagnetic dynamo which, according to a theorem by Cowling [8], cannot have a spherical flow pattern. To make an estimate about the magnitude of the energy loss due to gravitational radiation, we may take the expression derived by Eddington [9] for a spinning rod:

$$\mathbf{P} = \frac{32 G I_m^2 \omega^6}{5 c^5} \tag{13}$$

or as given by Weber [10] in cgs units

$$\mathbf{P} = 1.73 \times 10^{-59} I_m^2 \omega^6 \left[ erg/s \right] \tag{14}$$

where $I_m$ is the inertial momentum of the rod and ω its angular velocity. For $I_m$ we set $I_m = mr^2$, and for the angular velocity $\omega = v_o/r$. With m=6.8×10³²g, r=1.4×10¹⁰cm, $v_o$ =5.3×10⁷cm/s, one obtains

$$\mathbf{P} \simeq 10^{33} \, erg/s \tag{15}$$

At the distance of R=150 million km from the earth-moon system the intensity of gravitational radiation per cm² is reduced by 4πR²≃3×10²⁷cm² to the respectable value of

$$\mathbf{P}_E \simeq 3 \times 10^5 \, erg/cm^2 s \tag{16}$$

This large amount of gravitational radiation emitted by the sun, is only 5 times smaller than the solar constant (1.4×10⁶erg/cm²s) and comes as a surprise. The reason for it is the ω⁶ dependence of equation (13). However, it is still small enough to be in agreement with the age of main sequence stars and the universe, which would reduce the age by only 1/5th. Eddington's formula gives us only an estimate, **P,** and for the solar dynamo it could be much smaller. In theories of stellar structure [6], ignoring the possible existence of a powerful dynamo, the convection velocity is estimated to be of the order 10⁴cm/s, not the 10⁷cm/s. At these much smaller velocities there would be no appreciable emission of gravitational radiation.

## 4. Screening the Large Magnetic Field of the Dynamo Inside the Sun

For the dynamo theories of stellar and planetary magnetism the Coriolis force is of crucial importance. However, the assumed large solar dynamo leads to very large magnetic fields, not observed on the surface of the sun. This requires an explanation of how these fields are screened from reaching the surface.

The temperature dependence of the thermonuclear reaction rate in the region of $10^7$K goes in proportion to $T^{4.5}$ [6]. This means there is a sharp boundary between a much hotter region where most of the thermonuclear reactions occur and a cooler region where they are largely absent. This boundary is the tachocline. It is the thermomagnetic Nernst effect which by the large temperature gradient in the tachocline between the hotter and cooler region leads to large currents shielding the large magnetic field of the dynamo.

For a fully ionized hydrogen plasma the Nernst effect leads to a current density given by

$$\mathbf{j}_N = \frac{3knc}{2B^2} \mathbf{B} \times gradT \tag{17}$$

with the force density

$$\mathbf{f} = \frac{1}{c}\mathbf{j} \times \mathbf{B} = \frac{3}{2}\frac{nk}{B^2}(\mathbf{B} \times gradT) \times \mathbf{B} \tag{18}$$

or with gradT perpendicular to **B**

$$\mathbf{f} = \frac{3}{2}nk \times gradT \tag{19}$$

leading to the magnetic equilibrium condition

$$\mathbf{f} = gradp \tag{20}$$

with p=2nkT, one has gradp=2nk gradT+ 2kT gradn, hence

$$2nk \times gradT + 2kT \times gradn = \frac{3}{2}nk \times gradT \tag{21}$$

which upon integration yields

$$Tn^4 = const. \tag{22}$$

Taking a cartesian coordinate system with z directed along grad T, the magnetic field into the x-direction and the Nernst current into the y-direction, one has

$$j = j_y = -\frac{3knc}{2B}\frac{dT}{dz} \tag{23}$$

From Maxwell's equation 4πj/c=curl**B**, one has

$$j_y = \frac{c}{4\pi}\frac{dB}{dz} \tag{24}$$

and thus

$$2B\frac{dB}{dz} = -12\pi kn\frac{dT}{dz} \tag{25}$$

From (22) one has

$$n = \frac{n_o T_o^{\frac{1}{4}}}{T^{\frac{1}{4}}} \tag{26}$$

where for large values of z (towards the center of the sun) $n = n_o$, $T = T_o$. Inserting (26) into (25) one finds

$$dB^2 = -\frac{12\pi kn_o T_o^{\frac{1}{4}}}{T^{\frac{1}{4}}}dT \tag{27}$$

and hence

$$\frac{B_o^2}{8\pi} = 2n_o k T_o \qquad (28)$$

expressing the fact that the magnetic field of the thermomagnetic current in the tachocline neutralizes the magnetic field of the dynamo reaching the tacholine.

The thermonuclear energy releasing-dynamo generating the magnetic field is, of course, not subject to such a shielding effect. Some magnetic field can still leak out, but even this residual magnetic field is weakened by resistive heating in the solar plasma.

5. The Moon as a Gravitational Wave Antenna

It was Weber's idea [10] to use meter-size metallic cylinders with a large Q-value as an antennas to detect gravitational waves coming from space. His experiment failed because the sensitivity of his antennas was not large enough.

The situation is changed if one uses the moon as an antenna instead of a Weber bar. Not only is the mass of the moon much larger than the mass of a Weber bar, but it has a large Q-value, of about Q≈5000 [11]. With the moon as an antenna for gravitational waves from the sun, these waves are focused by Poisson diffraction into the center line of the lunar shadow during a total eclipse.
The scattering cross section S of the moon for gravitational waves is given by Weber [10]

$$S = \frac{15\pi G m Q (\beta r)^2}{8\omega c} \qquad (29)$$

where m=7.9×10²⁷g is the mass and r=1.74×10⁸cm is the radius of the moon. β≡ $k_r$ is the wave propagation vector of the gravitational wave, with β=2π/λ, where λ is the wavelength. Setting λ=2πr, one has βr=1. Making this assumption one finds that S=1.3×10¹⁷ cm², and $\sqrt{S} \simeq 2r$, which means that during a total eclipse the moon forms a large disk with regard to the incoming gravitational radiation with a radius equal to λ=2πr≈10⁹cm=10⁴km, about one order of magnitude smaller than the radius of the solar core. This disc leads to Poisson diffraction of the incoming radiation, focusing it onto the axis of the lunar shadow.

6. Connection to Observations

The intensity of gravitational waves is conveniently measured by the change in the distance dx between two fixed points, from dx to hdx during the passage of a gravitational wave. The value of h is determined by the equation [1]

$$ct^{03} = \frac{c^3 \omega^2 h^2}{8\pi G} \tag{30}$$

where $t^{03}$ is the 03-component of the Einstein pseudotensor, and where one has the set ($\mathbf{P}_E$ given by (16))

$$ct^{03} = \mathbf{P}_E \tag{31}$$

One obtains

$$h \approx 10^{-7} \tag{32}$$

a huge value in comparison to the sensitivity of a LIGO antenna of the order $h \lesssim 10^{-22}$. But it must be emphasized that the value $h \sim 10^{-7}$ was obtained as the largest possible, estimated using Eddington's formula. Because the fluid motion in a magnetohydrodynamic dynamo is not spherical symmetric, there can be no doubt that such a dynamo must be the source of gravitational waves, but the magnitude of it is uncertain, due to a lack of a solution of the dynamo equations, which only recently became possible with the advent of supercomputers. To explain the perturbation on a Foucault type paraconical pendulum as observed by Allais, Saxl, and Allen [3,4], a value much smaller than $h \sim 10^{-7}$ is sufficient, which is possible for a dynamo working with a much lower efficiency. These observations can by explained by setting $h \sim 10^{-10}$, which is smaller by the factor $10^3$. However, since the gravitational wave emission rate is, according to (30), proportional to $h^2$, this means that the solar dynamo only needs to have an efficiency of $\sim 10^{-6}$. Along with this much smaller needed efficiency B² would be reduced by the same factor, with B from $B \sim 10^8$ G to $B \sim 10^5$ G.

According to the general theory of relativity, for weak gravitational fields one has [12]

$$h \simeq 2\phi / c^2 \tag{33}$$

where Φ is the scalar gravitational field in Newtonian approximation, leading to the acceleration |a| given by

$$|a| = \nabla \phi \tag{34}$$

The gravitational waves emitted from the solar dynamo have a wavelength $\lambda = c / \omega \sim 10^{13} cm$ and for $h = 10^{-10}$, one has

$$a \sim hc^2 / \lambda \sim 10^{-2} cm/s^2 \tag{35}$$

During the one minute time of a total eclipse $t \sim 10^2$ sec, and for the displacement is $\delta \sim at^2 \sim 100$ cm, large enough to observe a substantial displacement of the pendulum.

Assuming that all the main sequence starts have dynamos, this would create a large stochastic background of gravitational waves, with a wavelength too long to be detected by LIGO antennas, but which might explain why no such signals can be detected by such antennas.

7. <u>Conclusion</u>

There can be no doubt that a stellar dynamo emits gravitational waves, the reason is that a dynamo is possible only with a non-spherical fluid flow. But because of the difficulty in obtaining solutions to the dynamo problem, only recently possible with the advent of supercomputers, little is known about the efficiency of such dynamos. However, even with the rather poor efficiency of $10^{-6}$, the observed strange disturbances of pendulums during a total solar eclipse can be qualitatively explained.

With the large wavelengths of several thousand kilometers, which cannot be observed by LIGO antennas, and with the assumed existence of solar star dynamos in all main sequence stars, this would create a large gravitational wave background which may black out the shorter wave signals which could otherwise be received with LIGO antennas and might explain why no signals have been detected.

References

1. F. Winterberg, Il Nuovo Cimento 53B, 264 (1968).
2. H. Alfvén, Cosmical Electrodynamics, Oxford University press, London, New York 1950.
3. Maurice F.C. Allais, Aerospace Engineering 18, 46 (1959).
4. E.J. Saxl and M. Allen, Physical Review. D3, 823 (1970).
5. L. Landau and E. Lifshitz, Electrodynamics of Continuous Media, Pergamon Press, Oxford 1960, p. 213 ff.
6. M. Schwarzschild, Structure and Evolution of the Starts, Princeton University Press, 1958, p. 44 ff.
7. John G. Ratcliffe, Foundations of Hyperbolic Manifolds, Springer Verlag, New York, Berlin 1994.
8. T. G. Cowling, Monthly Notices of the Royal Astronomical Society 94, 39 (1934).
9. A.S. Eddington, Proceedings of the Royal Society (London) A102, 268 (1923).
10. J. Weber, General Relativity and Gravitational Waves, Interscience Tracts on Physics and Astronomy, No.10, Interscience Publishers, Inc. New York, 1961.
11. Y. Nakamura and J. Koyama, Journal of Geophysical Research 87, 4855 (1982).
12. L. Landau and E. Lifshitz, The Classical Theory of Fields, Addison-Wesley Publishing Company, Massachusetts , USA, 1951, p. 327.